\documentclass[11pt,a4paper]{article}

\usepackage[cp1251]{inputenc}

\usepackage[russian]{babel}
\usepackage{amsmath}
\usepackage{amssymb,amsfonts,amsmath,mathtext}
\usepackage{cite,enumerate,float,indentfirst}
\usepackage{graphicx}
\usepackage{subfig}
\usepackage{xspace}

\newcommand{\bi}{\bibitem}
\newcommand{\be}{\begin{eqnarray}}
\newcommand{\ee}{\end{eqnarray}}
\newcommand{\rar}{\rightarrow}
\newcommand{\lrar}{\leftrightarrow}

\begin{document}

\title{\bf Термодинамика и кинетика элементарных частиц в космологии  }
\author{А.Д. Долгов }
\maketitle
\begin{center}
Новосибирский Государственный Университет, Новосибирск 630090\\
Институт Теоретической и Экспериментальной Физики, Москва 117218\\
Dipartimento di Fisica e Scienze della Terra, Universit\`a 
di Ferrara, I-44124 Ferrara, Italy \\
Istituto Nazionale di Fisica Nucleare, Sezione di Ferrara,
I-44124 Ferrara, Italy
\end{center}

\begin{abstract}
Обсуждается установление термодинамического равновесия в ранней вселенной, а также процессы, 
приводящие к отклонению от равновесия. Во вводной лекции кратко приведены основные сведения
из космологии. Далее рассматривается кинетическое уравнение в метрике Фридмана-Робертсона-Уокера.
Анализируется характер решений кинетического уравнения  для различных типов частиц, в частности,
фотонов микроволнового фона, нейтрино и массивных частиц, которые могли бы представлять собой
темную материю. Обсуждается модификация условия детального равновесия при нарушении 
Т-инвариантности, а также возможное измененение канонической кинетики при нарушении 
СРТ-теоремы и приложения к бариосинтезу. 

 \end{abstract}

\section{Введение} \label{s-intro} 

Как известно, на заре мироздания вещество во вселенной было горячим и очень плотным,
возникшем в результате так называемого большого взрыва (big bang).
Скорее всего, начальное состояние до большого взрыва представляло
собой экспоненциально расширяющуся квази-пустоту, заполненную вакуумо-подобным
состоянием инфлатонного поля. Впрочем, как говорят, тут возможны варианты,
и первоначальное экспоненциальное  расширение (инфляция)  могло бы идти за счет
каких-либо иных механизмов. Для нас будет важно, что в  какой-то момент эта
темная квази-пустота  (квази-вакуум) взорвалась с рождением большого числа всевозможных 
элементарных частиц.  

Поразительно, что внешне этот процесс очень напоминает библейское описание сотворения мира: "Земля была безвидна и пуста и тьма над 
бездною."  А потом: "Да будет свет!".  Однако, в отличие от Библии,  для описания 
процесса большого взрыва имеется строгая математическая теория, количественные 
следствия которой могут быть проверены астрономическими наблюдениями и, забегая
вперед, скажем, что предсказания теории замечательно согласуются с наблюдаемыми
чертами вселенной. 

Теория взрыва квази-вакуума или, иными словами, рождения частиц инфлатонным полем
хорошо установлена и позволяет рассчитать скорость рождения и спектр энергий возникших частиц. Этот спектр может иметь довольно сложный вид, но в результате
взаимодействия между рожденными частицами вся предудущая история оказывается
забытой и спектры всех (или почти всех) частиц приобретают универсальную форму,
отвечающую тепловому равновесию. В этом случае все разнообразие спектров  описывается 
всего лишь тремя параметрами: температурой, общей для всех частиц,  
химическими потенциалами, которые, вообще говоря, могут быть различны для 
разных типов частиц, и массами, которые различны для разных типов частиц.
Исключение представлает возможный Бозе конденсат в космологической плазме, когда
химический потенциал фиксирован и равен массе частиц в конденсате, но тем не менее,
распределение по-прежнему описывается тремя параметрами; вторым
параметром, вместо химического потенциала, является плотность или амплитуда 
конденсата (мы обсудим это ниже).

Интересно, что в обычных физических системах термодинамическое равновесие
достигается по прошествии достаточно длительного времени, а в космологии -
наоборот: равновесие существовало в ранней вселенной, а поздние отклонения от равновесия могли
достигать значительной величины. Эти отклонения возникают для массивных частиц, в то время 
как безмассовые, как правило, сохраняют равновесный спектр при космологическом расширении. 
Примером таких частиц с идеально (или
почти идеально) равновесным спектром являются фотоны космического
микроволнового фона (Cosmic Microwave Background radiation или CMB). 
Наблюдение этого излучения
в 1965 году Пензиасом и Вильсоном~\cite{penzias} поставило окончательную точку в споре между 
сторонниками горячей (big bang) и холодной вселенной.

Это одно из наиболее важных в космологии открытий следовало из предложенной Гамовым\cite{gamow} 
в 1946 году модели горячей вселенной. Конкретное предсказание существования фона микроволновых 
фотонов с температурой в несколько градусов Кельвина было сделано вскоре после этого
Альфером и Херманом~\cite{alpher-herman}.  В 1964 году Дорошкевич и Новиков~\cite{dor-nov}
показали, что интенсивность микроволнового фона в широкой области частот должна значительно 
превышать интенсивность электромагнитного излучения от других источников и его наблюдение
является вполне реалистичным. Однако, их работа,
по-видимому, не была известна Пензиасу и Вильсону. Интересно, что задолго до Пензиаса и Вильсона
в аналогичного типа эксперименте по калибровке антенны радиотелескопа Тер Шмаонов~\cite{shmaonov}
 обнаружил фон реликтовых фотонов, оценив его температуру в $4 \pm 3$ K.
В настоящее время изучение угловых и спектральных флуктуаций (т.е. отклонения от равновесного
распределения)  является одним из наиболее чувствительных способов измерения космологических 
параметров.

В этих лекциях мы рассмотрим процессы установления термодинамического равновесия и отклонения
от равновесия в ранней вселенной для фотонов, нейтрино, тяжелых частиц, которые могли бы быть
носителями темной материи. Во вводных разделах 2 и 3 обсуждаются необходимые космологические 
уравнения, включая уравнения Фридмана, а также приведены основные космологические параметры.
В разделе 4 рассмотрено кинетическое уравнение в расширяющйся вселенной и его решения в простых,
но практически интересных случаях. Также обсуждается условие циклического баланса, которое приходит
на смену условию детального баланса, если нарушена инвариантность относительно обращения
времени. В центральном разделе 5 рассмотрены процессы установления равновесия и отклонения
от равновесия для фотонов, нейтрино и массивных частиц. В разделе 6 обсуждается гипотетическая 
кинетика при нарушении СРТ-теоремы и возможный бариосинтез в таких условиях. В Приложениях 
обсуждаются термодинамические величины в равновесной плазме,
приведены примеры законов космологического расширения для различных уравнений состояния материи
и кратко описана тепловая история вселенной. Несколько более подробно 
разобран бариосинтез и первичный нуклеосинтез.

\section{Космологические уравнения} \label{s-cosm-eqs}

В этом разделе мы элементарно, без использования общей теории относительности (ОТО), хотя и не вполне строго, выведем основные космологические уравнения, т.е. уравнения Фридмана, и закон космологического
красного смещения, а также введем понятие уравнения состояния космологической материи.
Обсуждения этих вопросов можно 
найти в лекциях~\cite{lectures-cosm}, а строгий вывод уравнений во многих учебниках
по общей теории относительности, например в книге~\cite{landau}. Для более глубокого
ознакомления с предметом можно рекомендовать книги~\cite{text-books}.

Вскоре после создания ОТО, в 1922 году Фридман~\cite{friedman} нашел решения уравнений 
Эйнштейна для однородного и изотропного распределения вещества, что в хорошем приближении 
описывает вселенную на ранних стадиях ее эволюции. В работе Фридмана был предсказан 
удивительный
по тем временам результат, состоящий в том,
что вселенная не стационарна, она расширяется и была установлена
связь закона расширения с плотностью энергии материи. Установление космологического расширения
в астрономических наблюдениях
приписывается Хабблу~\cite{hubble}, хотя были и заметно более ранние работы, где это расширение
было открыто~\cite{pre-hubble}.

Для элементарного вывода уравнений Фридмана 
рассмотрим бесконечное пространство, заполненное веществом с однородной
плотностью энергии, которая может зависеть от времени, $\rho = \rho(t)$.
Выделим сферическую область радиуса $a(t)$ и
пробное тело на поверхности этой сферы. Энергия этого пробного тела
(кинетическая плюс потенциальная) должна сохраняться: 
\be 
\frac{\dot a^2}{2}=\frac{G_N M}{a}+const, \label{sokhr-E} 
\label{conserv-E}
\ee
где $G_N$ - ньютоновская гравитационная постоянная, а $M={4\pi a^3 \rho}/{3}$ - масса внутри 
шара радиуса $a(t)$. В естественной системе единиц, где $c=k=\hbar  =1$
можно записать $G_N = 1/m_{Pl}^2$, а масса Планка равна 
$m_{Pl} = 1.22 \cdot 10^{19}$ ГэВ. 

Отношение $H = \dot a/a$
называется параметром Хаббла и, если $H\neq 0$, то это означает, что вселенная 
расширяется, причем скорость убегания от нас отдаленного объекта пропорциональна расстоянию до него. 

Обычно уравнение (\ref{conserv-E}), первое уравнение Фридмана,  в космологии записывается в виде:
 \be
 H^2 \equiv \left(\frac{\dot a}{a}\right)^2 =
\frac{8\pi\,\rho\,G_N}{3} -
\frac{k}{a^2}\,,\;\;\;
\label{fridmaneq1}
\ee
где $k = const$, принимаемая равной $\pm 1$ или нулю. Выбор $k=\pm 1$ отвечает определенной
 нормировке масштабного фактора $a(t)$; при $k=0$ нормировка $a$ не фиксирована.  

Во избежание недоразумения заметим, что трехмерное расстояние между двумя точками не равно масштабному фактору $a(t)$, но пропорционально ему. Дело в том, что
фундаментальной величиной в ОТО является метрический тензор $g_{\mu\nu}$, который определяет интервал $ds$ согласно
\be 
ds^2 = g_{\mu\nu} dx^\mu dx^\nu .
\label{ds2}
\ee
Однородная и изотропная Вселенная описывается метрикой
Фридмана-Робертсона-Уокера (FRW) с интервалом, имеющим вид: 
\begin{equation}
ds^2 = dt^2 - a^2(t) \left[ dr^2 +f(r)
\left(d\theta^2 + \sin^2 \theta d\phi^2 \right)\right] ,
\label{ds2-fridman}
\end{equation}
где функция $f(r)$ зависит от топологии Вселенной в целом. Для
пространственно плоской Вселенной $f(r)= r^2$, для замкнутой
Вселенной $f(r)= \sin^2 r $ и для открытой Вселенной $f(r)= \sh^2r$, где функция времени $a(t)$ 
полностью определяется свойствами материи. Обычно, хотя и не всегда, свойства материи
задаются уравнением состояния, т.е. зависимостью плотности давления $ P $ 
от плотности энергии $\rho$.
 
Все  три указанных случая отвечают пространствам постоянной кривизны: 
замкнутой трехмерной сфере $(k=+1)$, открытому гиперболоиду $(k=-1)$ 
или трехмерному плоскому пространству нулевой кривизны $(k=0)$. 
Трехмерное расстояние при постоянных углах 
равно $dl^2= a^2 dr^2$,  где $r$ - так называемая сопутствующая координата, постоянная для каждой 
частицы.

Следующее необходимое нам уравнение -- баланс энергии материи. Так как 
$ dE=-P dV$ где $P$-плотность давления, то, подставляя $E=\rho V \sim \rho a^3$, получим 
соотношение:
\be
dE=Vd\rho+3\frac{da}{a}V\rho \label{dE} ,
\ee 
которое можно переписать в виде:
\be
\dot{\rho}+3H(\rho+P)=0.
\label{dot-rho}
\ee
Это уравнение является ковариантным законом сохранения тензора энергии-импульса в метрике (\ref{ds2-fridman}): $D_\mu T^\mu_\nu = 0$.

Дифференцируя уравнение (\ref{fridmaneq1}) и используя уравнение (\ref{dot-rho}), получим выражение для космологического ускорения:
\be
\frac{\ddot a}{a} = -\frac{4\pi\,G_N }{3}(\rho+3P)\,,
\label{fridmanequ2}
\ee
Хотя это уравнение и не является независимым, оно весьма удобно, т.к. напрямую выражает ускорение при расширении через плотности энергии и давления материи.

Итак, мы имеем два независимых уравнения для описания трех неизвестных функий $a(t)$, $\rho(t)$ и $P(t)$. Чтобы доопределить систему обычно вводят уравнение состояния вещества, $P = P(\rho)$. Во многих практически интересных случаях можно ограничиться линейным уравнением состояния: 
\be
P = w\rho,
\label{eq-of-state}
\ee
где $w$ - постоянный параметр, различный для разных форм материи;
$w=0$  для нерелятивистского вещества, $w=1/3$ для релятивистского газа и
$w=-1$ для вакуума или вакуумо-подобной материи.

Из-за космологического расширения импульсы всех свободных частиц в трехмерно плоской вселенной уменьшаются обратно пропорционально масштабному фактору.
Это явление красного смещения, по существу, представляет собой просто эффект Доплера в разбегающемся веществе. Рассмотрим изменение импульса свободно движущейся частицы с импульсом $p$, находящейся в некоторый момент времени
$t_0$ в точке с координатой $x$. При приходе этой частицы в точку $x+dx$ ее импульс, согласно преобразованию Лоренца,  изменится на величину $dp =  -du E $, где 
$du = H dx$ - относительная скорость точек $x$ и $x+dx$ вследствие космологического расширения. Т.к. $dx = v dt$, где $v$ - скорость частицы, а $p=vE$, то мы получим 
\be 
\dot p = - Hp.
\label{dot-p} 
\ee 
Это уравнение можно строго вывести из уравнения геодезической в пространстве с метрикой (\ref{ds2-fridman}). При этом, однако, если трехмерное пространство является 
искривленным, то помимо допплеровских членов, в правой части возникнут дополнительные слагаемые, зависящие от трехмерной кривизны.

Из уравнения (\ref{dot-p}) следует, что импульсы свободных частиц
при космологическом расширении убывают обратно пропорцианально
масштабному фактору: 
\be 
p\sim\frac{1}{a(t)}\sim\frac{1}{(z+1)},
\label{p-of-z} 
\ee
 где $z\equiv a(t)/a(t_0)-1$ называется космологическим красным смещением.

\section{ Основные космологические параметры}  \label{s-cosm-prmtr}

Как уже отмечалось, темп расширения вселенной определяется параметром Хаббла, который, 
в свою очередь, выражается через плотность энергии материи во вселенной
согласно уравнению (\ref{fridmaneq1}).  Его численное 
значение в современную космологическую эпоху составляет:
\be 
H = 100 h\, {\rm km/sec/Mpc},
\label{H-of-h}
\ee
где безразмерный параметр $h $ равен $h = 0.73 \pm 0.03 $. Соответственно 
современное значение характерного
космологического времени равно $H^{-1} \approx 4\cdot  10^{17} $ сек.
За исключением периода инфляции (т.е. первоначального экспоненциального расширения
вселенной), $H^{-1}$ примерно равняется возрасту вселенной, т.е. времени, прошедшему от 
момента большого взрыва. Соответственно, $H$ тем больше, чем ближе к
началу находилась вселенная. 

В пространственно плоской вселенной, т.е. при $k=0$ в уравнении (\ref{fridmaneq1}),
полная плотность энергии, называемая в этом случае
{\it критической плотностью}, однозначно выражается через параметр Хаббла:
\be 
\rho_c = \frac{3H^2\,m^2_{Pl}}{8\pi} \approx 10^{-29}\, {\rm g/cm}^3 \approx 5 \,{\rm keV/cm^3}.
\label{rho-c} 
\ee
Плотности энергии различных видов материи в космологии выражаются в терминах
$\rho_c$  в виде безразмерного отношения:
\be
\Omega_a  \equiv \rho_a/\rho_c. 
\label{Omega} 
\ee 
Значение полного параметра $\Omega_{tot} = \sum_a \Omega_a $ близко к единице:
\be 
\Omega_{tot} = 1.02 \pm 0.02
\label{Omega-tot} .
\ee
Иными словами, полная плотность энергии всех видoв материи близка к критической,
а, следовательно, геометрия вселенной в среднем эвклидова. В дальнейшем везде 
будем полагать $k=0$, если не оговорено противное. Более того, даже если в настоящее время
$\Omega_{tot}$ заметно  бы отклoнялась от единицы, тем не менее в ранней вселенной
это отклонение было бы ничтожно мало.

Обычное барионное вещество составляет лишь небольшую часть от полного:
\be
\Omega_B = 0.04 - 0.05.
\label{Omega-B}
\ee
Так называемая темная материя вносит вклад в 4-5 раз больше:
\be 
\Omega_{DM} \approx 0.2 
\label{Omega-DM} .
\ee
Уравнение состояния как барионного вещества, так и темной материи на достачно поздней стадии
 космологической эволюции, когда обе эти формы материи оказываются нерелятивистскими, описывается 
уравнением состояния с  $ w = 0$. 

Доминирующий вклад в полную плотность космологического  вещества вносит темная энергия с ''антигравитирующим''  уравнением состояния $w=-1$. Согласно уравнению (\ref{fridmanequ2}) 
такая связь между $P$ и $\rho$ приводит к ускоренному расширению вселенной. Вклад темной 
энергии составляет:
\be 
\Omega_{DE} \approx 0.75.
\label{Omega-DE}
\ee

Кроме указанных форм материи, во вселенной имеется почти однородный фон микроволновых фотонов (Cosmic Microwave Background Radiation или CMBR). Его вклад в полную плотность энергии невелик:
\be
\Omega_{\gamma} = 4.6\cdot 10^{-5},
\label{Omega-gamma}
\ee
но роль этого фонового излучения в космологии трудно переоценить. Спектр излучения
очень близок к планковскому, см.  ниже уравнение (\ref{f-eq}) для бозонов при $\mu =0$. 
Это говорит о том, что ранняя вселенная была горячей и состояние вещества было термодинамически
 равновесным. Температура CMB весьма аккуратно измерена и равна
\be 
T_0 = 2.725 \, {\rm K}.
\label{T-0}
\ee
Число этих фотонов в единице объема и их плотность энергии равны соответственно:
\be 
n_\gamma^{(0)} = 410.5/{\rm cm}^3,\,\,\, 
\rho_\gamma^{(0)} \approx 0.23\, {\rm eV/cm}^3.
\label{n-gamma-0}
\ee
Вывод этих соотношений и их выражения чeрез температуру приведены ниже в 
Приложении I.

Обычно при описании космологической истории используют не время, а красное смещение
$z$, определенное в уравнении (\ref{p-of-z}) и сразу же после него, причем $z=0$ 
отвечает настоящему времени, а более ранние моменты  - все большим значениям $z$. Если можно
пренебречь впрыском энергии в фотонную компоненту космологической плазмы, то $z$ растет
пропорционально температуре: $z(T) = T/T_0$.

Кроме фона микроволновых фотонов во вселенной должен быть также фон нейтрино с очень низкими
 импульсами, в районе $10^{-4}$ эВ. Этот фон непосредственно не обнаружен и вряд ли будет 
обнаружен в ближайшем будущем, но в его существовании никаких сомнений нет. Согласно 
оценкам $\Omega_\nu \sim  0.01-0.001$.
В отличие от фотонов спектр нейтрино неравновесный, т.к. они массивны. Кроме того
имеется поправка к спектру на процентном уровне из-за аннигиляции горячих электрон-позитронных пар в ранней вселенной, см. раздел. \ref{ss-nu}.

Полезно привести также отношение числа барионов к числу фотонов. Это отношение
сейчас очень надежно определяется по угловым флуктуациям микроволнового фона
и несколько менее аккуратно по обилиям легких элементов, главным образом дейтерия
и гелия - 4. Величина этого отношения равна:
\be
\beta = n_b/ n_\gamma \approx 6\cdot 10^{-10} .
\label{beta}
\ee

\section{Кинетическое уравнение } \label{s-kineq}

В приближении слабо-взаимoдействующего газа или плазмы имеет смысл говорить об одночастичной 
функции распределение частиц в фазовом пространстве $f(t, {\bf p,x})$. Средние характеристики системы 
выражаются через интегралы от этой функции распределения. Так например, плотность числа частиц в 
единице объема и их плотность энергии равны соответственно:
\be
n &=& \sum_s \int \frac{d^3 p}{(2\pi)^3} f ,
\label{n}\\
\rho &=&  \sum_s \int \frac{d^3 p}{(2\pi)^3} E(p) f .
\label{rho}
\ee
Здесь $E(p) = \sqrt{p^2 + m^2}$ - энергия свободной частицы с импульсом $p$ и массой
$m$; суммирование проводится по спиновым состояниям, т.е. по проекциям спина $s$.  
Возможна ситуация, 
когда функции распределения для разных спинов различны. В этом случае $f$  зависит 
еще и от $s$. Подобное положение имеет место, например, для нейтрино, когда в 
космологическом веществе обильно присутствуют лишь левые состояния, т.е. состяния с 
проекцией спина на направление движения (спиральностью) равной (-1/2) для нейтрино 
и (+1/2) для антинейтрино и практически отсутствуют правые состяния, т.е. состояния с 
противоположной спиральностью.

Выражения для равновесной плотности числа частиц в единице объема, их плотности 
энергии и другие полезные формулы приведены в Приложении I.

Функции распределения дла какого-то типа частиц $i$ удовлетворяют кинетическому 
уравнению, которое в общей форме записывается в виде:
\be
\frac{df_i}{dt}  = I_i^{(coll)},
\label{df-dt}
\ee
где $I^{(coll)}_i$ - интеграл столкновений для частицы $i$, а 
полная производная по времени расписывается следующим образом:
\be 
\frac{df}{dt} = \frac{\partial f}{\partial t} + {\bf \dot  p}\, \frac{\partial f}{\partial {\bf p}}
+ {\bf \dot x}\,\frac{\partial f}{\partial {\bf x}} .
\label{df-dt2}
\ee
Здесь ${\bf \dot x = v }$, где ${\bf v} $ - скорость частицы, которую следует выразить через ее импульс. В нерелятивистском пределе ${\bf v  =p}/m$. Используя уравнения движения, можно записать ${\bf \dot p = F}$, где ${\bf F}$ - внешняя сила, действующая на частицу.

Заметим, что переменные $t$, $ {\bf x}$ и ${\bf p}$, входящие в $f (t, {\bf p,x})$
 рассматриваются, как независимые, тогда как выше мы полагали ${\bf p = p} (t)$. 
В этом, однако, нет никакого внутреннего противоречия. Дело в том, что для
 определения эволюции $f$ мы должны рассматривать потоки физических частиц как в импульсном, 
так и в координатном пространстве, где они движутся согласно своим уравнениям движения.
Потому-то мы и используем $\dot p = -Hp$.

В однородной и изотропной вселенной функции распределения не зависят ни от координат, ни  от
 направления импульса, а  зависят лишь от времени и от абсолютной величины импульса или, что то 
же, от энергии. Используя уравнение (\ref{dot-p}), получим для левой части кинетического уравнения:
\be 
\frac{\partial f_i}{\partial t} -  Hp \frac{\partial f_i}{\partial  p} = I^{(coll)}_i
\label{lhs-cosm}
\ee
Интеграл столкновений для процесса $ i+Y \rar Z$, где $Y$ и $Z$ -  какие-то,  вообще 
говоря, многочастичные состояния, определяется следующим образом:
\begin{equation}
\label{I-coll}
I^{coll}_i=\frac{(2\pi)^4}{2E_i}\sum_{Z,Y}\int\,d\nu_Z\,d\nu_Y
\delta^4(p_i+P_Y-P_Z)\times
\end{equation}
$$ \times\left[|A(Z\to i+Y)|^2\prod_Z f\prod_{i+Y}(1\pm f) -
|A(i+Y\to Z)|^2f_i\prod_Y f\prod_Z(1\pm f)\right]. $$ 
Здесь $A(i+Y\to Z) $-амплитуда процесса перехода из состояния $i+Y$ в состояние $Z$,  
$\prod_Y f$ - произведение плотностей частиц, образующих состояние 
$Y$, знаки в $\prod(1\pm f)$
выбираются соответственно для бозонов или фермионов  и
\be d\nu_Y=\prod_Y\overline{dp}\equiv
\prod_Y\frac{d^3p}{(2\pi )^32E}\, .
\label{d-nu}
\ee 

Равновесными функциями распределения называются такие функции, которые обращают в нуль
интеграл столкновений. Можно проверить, что таковыми являются канонические 
распределения Бозе и Ферми: 
\begin{equation}
f^{(eq)}_{f,b}(E,T,\mu) =\frac{1}{\exp\left[{\left(E-\mu\right)}/{T}\right]\pm 1}.
\label{f-eq}
\end{equation}
Распределения зависят от двух произвольных параметров: температуры $T$ и
химического потенциала $\mu$, которые в космологической ситуации
являются функциями времени.

Утверждение, что $I^{(coll)} [ f^{(eq)} ] = 0$ легко проверить в случае, когда имеет место 
инвариантность относительно обращения времени, т.е. относительно Т-преобразования.
В этом случае абсолютная величина амплитуды обратного процесса равна величине 
амплитуды прямого с  Т-преобразованными значениями импульсов и спинов, 
т.е. при  $p \rar (-p)$ и $s \rar (-s)$.   
Делая соответствующую замену в интеграле столкновений, отвечающему 
обратному процессу, мы увидим, что амплитуда не изменится и поэтому
можно вынести квадрат ампитуды за скобки, в результате чего  
под интегралом возникнет множитель, пропорциональный разности:
\be 
\prod_Z f\prod_{i+Y}(1\pm f) - f_i\prod_Y f\prod_Z(1\pm f).
\label{diff}
\ee
Простой алгеброй можно убедиться, что эта разность обращается в нуль в силу того, что
сохраняются энергии и химические потенциалы,
т.е. их суммы в начальном и конечном состояниях равны:
\be
\sum \mu_{in} =\sum \mu_{fin}.
\label{chem-pot-equil}
\ee
Условие сохранения энергии справедливо всегда, когда нет внешних полей, зависящих 
от времени, а сохранение химических 
потенциалов имеет место только в равновесной ситуации. Система сама приходит в это 
состояние в процессе эволюции, если скорость реакций достаточно велика. 

Однако следует помнить, что при упругом рассеянии условие сохранения 
химического потенциала выполняется автоматически. Поэтому упругие
реакции приводят систему в состояние кинетического равновесия, формируя 
каноническую зависимость функций распределения от энергии, но ничего не делают с
химическими потенциалами, оставляя их произвольными, точнее, определяемыми
начальными условиями.

Рассмотрим процесс аннигиляции электрон-позитронной пары в два и три фотона:
\be 
e^+e^- \leftrightarrow 2\gamma,\,\,\,  e^+e^- \leftrightarrow 3\gamma
\label{ee-gamma}
\ee
Условие равновесия для этих реакций гласит:
\be 
\mu_{e^-} + \mu_{e^+} = 2\mu_\gamma = 3 \mu_\gamma\, .
\label{mue-mugamma}
\ee
Отсюда сразу следут, что в равновесии
\be
\mu_\gamma = 0,\,\,\,\,\,\,\,\, \mu_{e^+} = -\mu_{e^-}\,.
\label{mu-gamma}
\ee

Измерения частотного спектра микроволнового фона показывает, что 
$\mu_\gamma = 0$, точнее $\mu_\gamma/T_\gamma < 10^{-4}$,
что является важным аргументом в пользу термодинамического
равновесия в ранней вселенной.

Второе условие является частным случаем общего утверждения, что в равновесии
химические потенциалы частиц и античастиц равны по величине и противоположны по
знаку.

Заметим, что вероятность реакции аннигиляции в три фотона примерно на 4 порядка 
меньше, чем вероятность, двухфотонной аннигиляции. Такое же соотношение справедливо и 
для  процессов
упругого $e\gamma$-рассеяния и тормозного излучения $\gamma e \rar 2\gamma\,e$. 
Поэтому может возникать (интересная для космологии) ситуация, когда, например, 
распределение фотонов имеет обычную зависимость от энергии (\ref{f-eq}), а их 
химический потенциал не равен нулю. Наблюдаемая малость $\mu_\gamma$ 
говорит о том, что  реализуется полное равновесие, как кинетическое, так и химическое
в полном соответствии с кинетической теорией в ранней вселенной.

При нулевом химическом потенциале плотности частиц и античастиц будут равны, как
это видно из формулы (\ref{f-eq}) и равенства масс частиц и античастиц. Последнее 
является следствием СРТ- теоремы и в случае нарушения СРТ-инвариантности плотности частиц
и античастиц могут различаться даже в полном равновесии при нулевых химических 
потенциалах и, наоборот, чтобы обеспечить равенство плотностей частиц и античастиц
надо ввести ненулевые химические потенциалы.
В нормальной же теории, где имеет место  СРТ-инвариантность, необходимо
ввести химический потенциал, чтобы описать ситуацию, когда плотности частиц и 
античастиц различны.  В таком случае, интегрируя выражение (\ref{f-eq}), например, 
для фермионов, получим
для разности плотности чисел безмассовых частиц и античастиц в единице 
обьема \footnote{ Вычисление интегралов от равновесных функций можно найти,
например, в книге~\cite{LL-kinetics}.}:
\be
 n -\bar  n =g_s\,T^3\,\frac{\xi^3 +\pi^2 \xi}{6\pi^2}
\label{n-barn}
\ee
где $g_s$ - число спиновых состояний, а $\xi = \mu/T$.

Увеличивая $\mu$,  можно увеличить асимметрию между частицами и античастицами
до любой, как угодно большой величины. Однако, это верно только для фермионов.
Для бозонов химический потенциал ограничен сверху массой соответствующего
бозона, $\mu < m$. Это условие очевидно, т.к. в противном случае функция  
распределения, $f$, может стать отрицательной для малых импульсов, что физически
бессмысленно. Это разумеется не означает, что асимметрия между бозонами и
антибозонами не может стать произвольно большой. Решение проблемы состоит в том, что при $\mu = m$ и только при $\mu = m$ возникает дополнительная однопараметрическая  свобода, которая позволяет равновесной функции распределения
иметь  вид:
\be
f^{(eq)}_B  (E,T, m,C)=\frac{1}{
\exp\left[{\left(E-m\right)}/{T}\right] - 1} + \frac{C}{(2\pi)^3}\, \delta^{(3)} ({\bf p}).
\label{f-eq-B}
\ee
Нетрудно проверить, что эта  фукция действительно зануляет интеграл столкновений,
т.е. является равновесной.  Слагаемое, пропорциональное дельта-функции от импульса, 
представляет собой Бозе конденсат, т.е. ансамбль частиц, накопившихся в состоянии с 
нулевым импульсом. Функция распределения по-прежнему описывается двумя 
параметрами, но теперь, вместо $T$ и $\mu$, мы имеем $T$ и амплитуду 
конденсата $C$. 

Для доказательства того, что распределения (\ref{f-eq}) или (\ref{f-eq-B}) 
обращают в  нуль интеграл столкновений, мы 
использовали предположение о $T$-инвариантности. Однако известно, что
$T$-инвариантность нарушается и возникает вопрос, сохранится ли форма равновесных 
распределений и в $T$ неинвариантной теории? Рассмотрим интеграл столкновений,
не предполагая T-инвариантности, подставим в него равновесные функции распределения и проверим, занулится ли он. Учитывая, что разность (\ref{diff})
равна нулю, вынесем за скобки произведение $\prod_Z f\prod_{i+Y}(1\pm f) $ 
и получим, что подынтегральное выражение имеет вид:
\be 
 \label{cyclebal}
\Delta \equiv \sum_k\int\,d\nu_k\delta^4(p_k-p_i)\Pi_k (1\pm f_k)
\left(|A_{ki}|^2-|A_{ik}|^2\right).
\ee
Последний сомножитель в круглых скобках, вообще говоря, отличен от нуля, если нарушается
Т-инвариантность. Тем не менее, можно показать, что хотя каждое отдельное слагаемое
в сумме (\ref{cyclebal}) отлично от нуля,  сумма в целом обращается в 
нуль~\cite{ad-cycle} в силу условия унитарности S-матрицы. Тут можно было бы 
возразить, что в принципе допустима ситуация, когда открыт лишь один канал
реакции и тогда в сумме (\ref{cyclebal}) присутствует только один член. 
Однако, хотя $\left(A_{ki}-A_{ik}\right) \neq 0$ при нарушении Т-инвариантности, 
весь эффект этого нарушения сводится к разности фаз амплитуд при равном модуле,
так что разность модулей амплитуд по-прежнему равна нулю.

Можно показать, что вместо условия унитарности достаточно
потребовать выполнения двух более слабых условий, а именно, $CPT$-инвариантности и 
сохранения вероятности:
\be
\sum_fw_{if}=1.
\label{cons-prob}
\ee
Если нарушается $CPT$, то, тем не менее, равновесная статистика останется 
невозмущенной, если выполняются соотношения:
\be
\sum_fw_{if}=1\,\,\,\mbox{и}\,\,\, \sum_fw_{fi}=1.
\ee
Если бы оказались нарушенными и $CPT$-инвариантность, и унитарность,
то в такой патологической (или революционной теории) могли бы
 возникнуть заметные макроскопические эффекты как в кинетике, так и 
в равновесной статистике.

Резюмируя, отметим, что справедливость обычной равновесной статистики практически
невозможно нарушить. В старые времена, когда Т-инвариантность не подвергалась
сомнению, вывод о канонической форме равновесных распределений основывался
на справедливости условия {\it детального баланса}, когда вероярности каждого
отдельного прямого и обратного процессов оказывались равными. В мире, в котором
мы живем, Т-инвариантность нарушена, но равновесная статистика не меняется
в силу более слабого условия {\it циклического баланса}~\cite{ad-cycle}, который
требует равенства суммы вероятностей всего цикла прямых и обратных реакций, хотя  
детальное равенство для каждой отдельной реакции может и не выполнятся.

\section{ Термодинамическое равновесие в ранней вселенной} \label{s-equil}

\subsection{Общие соображения} \label{ss-general}

Весьма важным и сильно упрощающим теоретический анализ космологии ранней 
вселенной является то обстоятельство, что первичная плазма находится в состоянии
очень близком к термодинамически равновесному. В каком-то смысле эта ситуация
противоположна установлению равновесия в обычных физических системах: 
стационарная замкнутая система приходит в состояние равновесия по истечении
достаточно длительного времени. Во вселенной же равновесие устанавливается на ранней
стадии, т.е. при коротких временах. Дело в том, что скорость реакций пропорциональна
плотности частиц, которая очень велика на ранней стадии. Причем этот эффект сильнее,
чем уменьшение существенного временного промежутка $t_{eff} \sim 1/H$. Исключение представляет 
распад частиц, т.к. время жизни частицы не зависит от плотности (если отвлечься от
влияния среды, т.е. от Ферми-запрета или Бозе-усиления).

Вселенная, как мы знаем, нестационарна, она расширяется и за счет этого температура 
падает, примерно как обратный масштабный фактор, если пренебречь выделением
энтропии, см. конец Приложения II. Предполагая справедливость гипотезы о 
термодинамическом равновесии, оценим темп расширения мира, сравним его со
скоростью реакций и увидим, что действительно скорость реакций намного больше
темпа расширения. Потому-то и возникает тепловое равновесие.

Темп космологического расширения определяется параметром Хаббла, который 
выражается через плотность энергии первичной плазмы согласно 
соотношению (\ref{fridmaneq1}), где мы положим $k =0$. Как легко понять, в плотности
энергии равновесной первичной плазмы доминируют релятивистские частицы, т.к.
вклад тяжелых частиц экспоненциально подавлен, см. уравнение (\ref{n-b-f-m}).
В силу этого согласно соотношениям (\ref{P-f-b}) и (\ref{rho-b-0}) получим:
\be 
\rho (T) = \frac{\pi^2 g_* (T)}{30}\, T^4.
\label{rho-of-T}
\ee  
Здесь $g_* (T)$ - количество типов элементарных частиц с учетом числа их спиновых состояний, вносящих вклад в полную плотность энергии: $g_* = g_b + (7/8) g_f $,
где $g_{b,f}$ - вклады бозонов и фермионов. К примеру при температуре плазмы
от $T = m_e = 0.511$ МэВ до, скажем, 50 МэВ ней присутствуют фотоны (дают 
$g_\gamma = 2$), электрон-позитронные пары (дают $g_e = 7/2$) и три типа 
лево-поляризованных  нейтрино (дают $ g_\nu = 21/4 $). Все вместе это сотавляет:
$g_* = 10.75$. Очевидно, $g_*$  зависит от температуры, т.к. при большей температуре
все больше типов частиц с $m<T$ заселяют первичную плазму. 

Учитывая вышесказанное, найдем:
\be
H = 5.44\sqrt{\frac{g_* (T)}{10.75}}\,\frac{T^2}{ m_{Pl}}\,.
\label{H-of-T}
\ee
Из-за присутствия большого фактора в знаменателе, $1/m_{Pl}$, темп расширения
оказывается, как правило, намного ниже скорости реакций между частицами, которые
имеют типичную скорость порядка:
\be
\Gamma \sim \sigma v n \sim \alpha^2 T,
\label{Gamma}
\ee
где $\sigma \sim \alpha^2/T^2$ -сечение взаимодействия частиц, $v$ - их скорость
и $\alpha \sim 10^{-2}$ -характерная величина константы взаимодействия. Очевидно,
равновесие установится при $ T < m_{Pl}/\alpha^2 \sim 10^{15}$ ГэВ.
Равновесие нарушается также и при низких температурах, во-первых, из-за того, что  
сечение может падать с темпeратурой, как, например, сечение слабого взаимодействия, 
$\sigma_W \sim G_F^2 T^2$. Этот пример мы рассмотрим ниже при обсуждении выхода
нейтрино из равновесия. Во-вторых, при $m>T$ плотность таких тяжелых частиц резко
падает и они выходят из равновесия. С этим связано, в частности, рассмотренное ниже 
явление замерзания (термин в англоязычной литературе) или, что то же, закалки 
(оригинальный термин, предложенный Зельдовичем) концентрации.

\subsection{Равновесие безмассовых частиц. Фотоны.} \label{ss-massless}

Представляется очевидным, что при большой скорости реакций распределение
частиц будет близким к равновесному. Однако мы знаем, что фотоны микроволнового
фона имеют равновесный планковский спектр с очень хорошей точностью, хотя
 их взаимодействие выключилось давным давно после рекомбинации водорода
(см. ниже в этом разделе). Оказывается, безмассовые частицы сохраняют свое исходно 
равновесное распределение, несмотря на расширение вселенной, даже когда 
выключено взаимодействие. Чтобы просто увидеть это, переопределим переменные в 
кинетическом уравнении: $x = m_0 a$ и $y_i = p_i a$, где $m_0$ - какая-то 
нормировочная масса, а  $p_i$ - импульс  частицы $i$. Тогда уравнение примет вид:
\be 
Hx\frac{\partial f_i}{\partial x} = I^{coll}_i\,. 
\label{df-dx}
\ee
При выключении взаимодействия, когда $I^{(coll)} = 0$, решение имеет вид
\be
f = f (y,x_{in}),
\label{f-of-y}
\ee
т.е. если в начальный момент $x_{in}$ распределение имело равновесную форму, оно ее 
сохранит и при выключении взаимодействия.

Большой интерес представляет промежуточная ситуация, на достаточно поздней стадии
космологической эволюции,
когда процесс упругого рассеяния $\gamma e\lrar  \gamma e$ был еще быстрым по
сравнению с темпом космологического расширения, но неупругая реакция
$\gamma e \lrar 2\gamma e$ уже вышла из равновесия. В стандартном космологическом
сценарии фотоны сохранят свое первоначально равновесное распределение. Однако 
возможны модификации стандартной модели, когда  в результате аннигиляции или 
распада каких-то новых массивных частиц произойдет впрыск энергии в фотонную компоненту, 
причем исходно спектр этих новых фотонов может сильно отличаться от равновесного.
Очевидно, что быстрое упругое рассеяние приведет к термализации фотонов, т.е. приведет к
каноническому распределению по энергии (\ref{f-eq}), но химический потенциал,
эволюция которого определяется гораздо более медленными неупругими процессами, не
успеет обратиться в нуль. 

Оценим характерное время и тех, и других процессов установления кинетического и 
химического равновесия. 
Фотоны находятся в равновесии с электронами вплоть до рекомбинации водорода, 
после которой электроны соединяются с протонами, образуя нейтральный водород
или с ядрами гелия, образую нейтральные атомы гелия. Это происходит примерно
при температуре 3000 К. Характерное время упругого $\gamma e$-взаимодействия
равно
\be
\tau_{\gamma e} = \left(c \sigma_{Th} n_e \right)^{-1},
\label{tau-gamma-e}
\ee 
где $ c $ - скорость света, принимаемая равной единице, 
$\sigma_{Th} = 8\pi \alpha^2/3m_e^2 = 6.7\cdot 10^{-25}$ см$^2$ - томсоновское сечение 
и  $n_e$ - число свободных электронов в единице объема. При $T<m_e$, но
выше рекомбинции, когда $e^+e^-$-пары уже аннигилировали, 
плотность электронов определяется
барионной асимметрией вселенной (\ref{beta}) и равна:
$ n_e \approx 1.5 \cdot 10^{-10} T^3 $. Соответственно
$\tau_{\gamma e} \approx 2\cdot 10^{12}\,(3000\,{\rm K}/T)^3 $ сек,
что заметно короче космологического времени вблизи рекомбинации,
$H^{-1} \sim 10^{13} $ сек. При последней оценке мы полагали, что во вселенной
доминирует нерелятивистское вещество и, следовательно, 
$ H \sim \sqrt{\rho} \sim a^{-3/2} $, где масштабный фактор убывает от нашего времени
до рекомбинации пропорционально отношению температур $ 3000/2.7 = 1100 $.    
После рекомбинации плотность свободных электронов падает, примерно, на 5 
порядков и взаимодействием фотонов микроволнового фона с космическими электронами
можно пренебречь. В силу этого частотный спектр фотонов не меняется, оставаясь 
равновесным, если он исходно был таковым. 

Приведенный результат для $\tau_{\gamma e}$ является хорошим приближением для оценки длины
 свободного пробега фотонов при рассеянии на нерелятивистских электронах, т.к. направление импульса
фотона может значительно измениться даже при однократном рассеянии. 
Что же касается возмущений в частотном спектре фотонов, то характерное время их замывания 
оказывается значительно дольше. Дело в том, что величина
импульса или, что то же, частота фотона при однократном рассеянии меняется очень мало:
\be
\delta \omega /\omega \sim v_e \sim \sqrt{T / m_e},
\label{delta-omega}
\ee
где $v_e$-скорость электрона и мы предположили, что электроны имеют тепловое распределение
с температурой $T$. Поэтому для того, чтобы относительное изменение частоты фотона оказалось 
порядка единицы, было бы необходимо $\sqrt{m_e/T}$ рассеяний, если бы они действовали однонаправлено.
Но так как этот процесс стохастический, как броуновское движение, то частота фотона при столкновениях
будет расти, как квадратный корень из числа столкновений и восстановление частотного спектра 
потребует $m_e/T$ столкновений. Соответственно необходимое время термализации возрастет до
$\tau_{therm} \approx 10^{16} ({\rm eV}/T)^4 $ сек. Оно сравняется с космологическим временем при
$ T \sim 10^2$ эВ, т.е. при красном смещении $z_{therm} \sim 5\cdot 10^5 $, где для космологического 
времени мы использовали оценку   (\ref{t-of-T}).

Оценим еще характерное время неупругой реакции $\gamma e \lrar 2\gamma e$, которая может влиять на 
химический потенциал фотонов, устремляя его к нулю, если бы он исходно был отличен от нуля. 
Сечение этого процесса при малой скорости электронов можно записать в виде~\cite{LL4}:
\be
d\sigma (\gamma e \rar 2\gamma e)
\approx d\sigma_{Th} \frac{\alpha d\omega}{4\pi^2 \omega}\, d o
\left( {\bf v' \times n} -   {\bf v \times n} \right),
\label{sigma-2gamma}
\ee
где $do $-элемент телесного угла,
${\bf v}$ и ${\bf v'}$ - скорости начального и конечного электронов, 
${\bf n}$ -  единичный вектор в направлении импульса фотона, а $\omega$ - частота
фотона. Для оценки по порядку величины положим $ v\sim v' \sim \sqrt{T/m_e}$, 
интеграл по $d{\bf n}$ равным $4\pi$ и интеграл $d\omega/ (\pi \omega) =  1$.
В результате для времени реакции тормозного излучения найдем:
\be
\tau_{2\gamma e} \approx \tau_{\gamma e} \left( \alpha \sqrt{m_e/T} \right)^{-1} 
\approx 10^5 \tau_{\gamma e} \left( T/{\rm eV}\right)^{1/2}.
\label{tau-2gamma}
\ee
Это время окажется короче космологического при $T> 10^7 $ эВ.  Однако температура, при которой
процесс тормозного излучения становится равновесным, значительно ниже. Дело в том,
что мы не учли, что при высоких температурах плотность $e^+ e^-$-пар перестает быть
пренебрежимо малой и окажется выше принятого нами значения (\ref{beta}). Используя
выражение (\ref{n-b-f-m}), найдем, что тормозное излучение могло эффективно изменять
химический потенциал фотонов при $T > 30$ кэВ или при красном смещении порядка $10^8$. 
Более точные вычисления приводят к насколько более низкому 
значению температуры, см., например, книги~\cite{zeld-nov}.

Таким образом, если неравновесные фотоны впрыскиваются в плазму при 
$ 10^2\,{\rm eV} < T <  3\cdot 10^4\,{\rm eV}$, то упругое 
$\gamma e$-рассеяние должно бы привести
функцию распределения фотонов к виду (\ref{f-eq}) c $\mu \neq 0$. Т.к. масса фотона равна 
нулю, то $\mu$ может быть только отрицательным. Однако требуемое значение $\mu$ зависит
от соотношения плотности энергии, $\delta \rho$, 
и числа дополнительных фотонов, $\delta n$, в единице объема.
Условие $\mu < 0$ будет выполнено лишь при достаточно большой плотности $\delta \rho$
по сравнению с $\delta n$. Мы можем легко оценить требуемое соотношение, пренебрегая 
космологическим расширением, хотя учет последнего и не приводит к большим трудностям.

В результате термализации за счет упругого рассеяния температура фотонов станет 
$T' = T +\delta T$, где  $T$-начальная температура равновесной части фотонного газа.
Полагая, что $\delta T /T $ и $\mu/T$ малы, разложим выражения (\ref{n}) и (\ref{rho}) до первого
порядка по $\xi\equiv \mu/T$:
\be 
n(T,\mu) &=& \frac{T^3}{\pi^2} \int \frac{dz z^2}{\exp (z) -1} \left( 1 + \frac{\xi}{\exp (z) -1}\right),
\label{n-of-mu}\\
\rho (T,\mu) &=& \frac{T^4}{\pi^2} \int \frac{dz z^3}{\exp (z) -1} \left( 1 + \frac{\xi}{\exp (z) -1}\right).
\label{rho-of-T-mu}
\ee
Сравнение численно вычисленных точных интегралов с этим приближенным ответом показывает
очень хорошее согласие при малых $\xi$. Например, при $\xi =-0.2$ отличие составляет 7\%
 для $n$ и 3\%  для $\rho$.

В силу того, что при упругом рассеянии число фотонов и их
плотность энергии сохраняются, мы можем записать:
\be
n(T',\mu ) = n(T,0) + \delta n,\,\,\,\, \rho(T',\mu) = \rho (T,0) + \delta \rho,
\label{n-of-t'-mu}
\ee
где $n(T,0) = 0.24 T^3$ и $\rho (T,0) = \pi^2 T^4/15$.

Какая-то перекачка энергии возможна в электроны, но их число очень мало по сравнению
с числом фотонов, а кроме того, как мы увидим, проблема возникает при малой величине 
$\delta \rho$, поэтому дополнительная утечка энергии в электронный сектор только усугубит 
проблему. 

Входящие в выражения (\ref{n-of-mu}) и (\ref{rho-of-T-mu}) интегралы при $\xi=0$ известны,
а те, что пропорциональны $\xi$, можно взять численно
и мы получим следующие выражения для $\delta T/T$  и $\mu/T$ через 
$\delta n \equiv 0.24 T^3 x$ и $\delta \rho \equiv (\pi^2/15) T^4 y$, где мы ввели новые безразмерные 
параметры $x$ и $y$. Соответственно получим:
\be 
\xi \equiv \mu/T \approx 3.5 x - 2.6 y,\,\,\,\, 
\delta T /T \approx  0.3 y - 0.1 x.
\label{xi-delta-T}
\ee
Видно, что $\mu$ имеет разрешенное отрицательное значение лишь при достаточно большом
впрыске энергии по сравнению с числом фотонов, $y > 4 x/3$. 

Возникает естественный вопрос, а что же будет, если это соотношение не выполняется? по-видимому,
ответ состоит в том, что предположение о сохранении числа фотонов не справедливо. И это 
действительно так, потому что каждый процесс упругого рассеяния сопровождается классическим
излучением большого числа мягких фотонов, которые термализуются, приобретая тепловой спектр 
при  последующем рассеянии на электронах. Однако, согласно приведенным выше соотношениям
проблема возникает не из-за недостатка фотонов, а  из-за их избытка. Поэтому необходим механизм
поглощения лишних фотонов. В принципе он мог бы осуществляться за счет обратного тормозного
излучения, т.е. поглощения магких фотонов в реакции $2\gamma e \rar \gamma e$.

Спектр реликтовых фотонов мог бы быть искажен из-за появления в плазме
более горячих или более холодных электронов. Это могло бы произойти в результате   
распада каких-то новых долгоживущих частиц с испусканием электрона или аннигиляции легких
частиц в $e^+e^-$ пары. Интересным кандидатом являются милли-заряженные частицы с массой
порядка МэВ~\cite{bdt}. Другим источником неравновесных электронов
мог бы быть тяжелый заряженный бозон, распадающийся на электрон и тяжелое нейтрино. 
Последнее предположение
не слишком реалистично, но может подойти, как модель для рассмотрения интересной кинетики. 
Аналогичное явление имеет место, когда реликтовое излучение попадает в область, где имеется
ионизованное вещество. При этом возникает известный эффект возмущения спектра, найденный
Зельдовичем и Сюняевым~\cite{zs} при расеянии реликтового излучения на горячих электронах
в галактиках или их скоплениях. При этом спектр фотонов отклоняется от равновесного
спектра и не описывается температурной функцией (\ref{f-eq}). При малых частотах возникает
дефицит фотонов, а при больших - избыток. Это связано с тем, что время взаимодействия (или
оптическая толща)  недостаточно велико, чтобы фотоны термализовались. При полной упругой
термализации температура фотонов должна была бы вырасти. Если бы их число не менялось, как
это обычно предполагается, то такое состояние было бы можно описать формулой (\ref{f-eq}) 
с $\mu<0$. Однако при рассеянии на холодных электронах с $T_e < T_\gamma$ полная 
термализация потребовала бы положительного химического потенциала, что невозможно. Это
указывает, что предположение о постоянстве числа фотонов должно быть отвергнуто и в этой ситуации.
Однако, т.к. при однократном рассеянии передача энергии от электрона к мягкому фотону
составляет всего $\delta \omega_{BS} /\omega_{BS} \sim \sqrt{T_e/m_e}$, то термализация тормозных 
(BS=bremsstrahlung) фотонов потребует очень большого числа соударений: 
$N\sim \sqrt{m_e/T_e}(T_\gamma/\omega_{BS})$ и, следовательно, большого времени, поэтому
в реалистической ситуации спектр мягких фотонов может заметно отличаться от равновесного.

\subsection{Неравновесность массивных частиц} \label{ss-massive}

Массивные частицы в космологии всегда отклоняются от равновесия. Это отклонение
может быть совсем небольшим или значительным и в последнем случае играет важную
роль в формировании современной вселенной. В частности, в обычном сценарии с
зарядово симметричной вселенной холодная или теплая темная материя практически бы
отсутствовала, если бы она когда-то не вышла из равновесия.

Для оценки величины отклонения от равновесности на ранней стадии запишем кинетическое 
уравнение (\ref{df-dx}) в виде:
\be
Hx \partial_x f = -\Gamma (f - f^{(eq)}),
\label{Gamma-delta-f}
\ee
где интеграл столкновений записан в приближенном линеаризованном виде,
а $\Gamma$ представляет собой эффективную скорость взаимодействия.
Например, для распада частицы $\Gamma$ - это ширина
распада, а для рассеяния $\Gamma = v\sigma n$, где $\sigma$ - сечение процесса,  $n$ - плотность 
сталкивающихся частиц, а $v$ - их скорость в системе центра масс. 

Предполагая, что отклонение от равновесия $\delta f = f- f^{(eq)}$ невелико, получим
 \be
\delta f \approx (H/\Gamma)\, x\partial_x f^{(eq)}.
\label{delta-f}
\ee
Обычно заметное отклонение от равновесия возникает при температурах сравнимых с массой.
Напомним, что для безмассовых частиц равновесие не нарушается. Посему положим $m>T$ и будем
использовать статистику Больцмана с $f^{(eq)} = \exp (-\sqrt{x^2+y^2}) $. В этом случае
\be
\delta f = \frac{H}{\Gamma}\, \frac{x^2}{\sqrt{x^2+y^2}} \approx \frac{Hx}{\Gamma} 
\sim \frac{mT}{m_{Pl} \Gamma}.
\label{delta-f2}
\ee
Здесь по порядку величины либо $\Gamma \sim \alpha m$ (для распада), либо
$\Gamma \sim \alpha^2 m$ для двухчастичного рассеяния и 
$\alpha \sim 10^{-2}$ - типичная величина константы взаимодействия, хотя она может быть
и намного меньше. При оценке сечения рассеяния предполагалось, что $T\sim m$.

Приведенные здесь общие оценки могут модифицироваться в разных физических условиях.
Например, равновесие относительно аннигиляции массивных частиц, обычно нарушается
при $T \sim 0.1 m - 0.01 m$, однако их кинетическое равновесие может поддерживаться
намного дольше (до более низких $T$) за счет упругого рассеяния с доминирующими
легкими частицами в плазме. Дело  в том, что взаимодействие с легкими частицами 
пропорционально плотности последних, которая порядка $T^3$, в то время, как вероятность
аннигиляции пропорциональна плотности партнеров для аннигиляции, которая при 
$T<m$ экспонениально подавлена. Совершить парное "самоубийство" становится все труднее
с падением температуры и потому частицы темной материи могут выжить в космологически
интересном количестве (см. ниже раздел \ref{ss-freezing}).

Если тяжелые частицы находятся в кинетическом равновесии с безмассовыми, которые доминируют
в первичной плазме, то их температура оказывается равной температуре безмассовых частиц и
падает по закону $T\sim 1/a$. Если же массивные частицы живут сами по себе во взаимном 
кинетическом равновесии, то их распределение должно иметь форму $f\sim \exp(-p^2/2mT)$.
При этом в силу того, что $p$ падает, как $1/a$, температура должна падать, как $1/a^2$. Контактируя
с другими легкими частицами, эти массивные частицы работают, как "холодильник'', приводя к более
быстрому остыванию легких частиц. 

Если же массивные частицы полностью прекратили всякое взаимодействие, то их функция
распределения может напоминать равновесную, только вместо энергии частиц она будет,
согласно уравнению  (\ref{f-of-y}),
зависеть от их импульса, как это имеет место для массивных нейтрино в современной вселенной, 
см. ниже уравнение (\ref{f-nu-now}).

\subsection{Неравновесность нейтрино} \label{ss-nu}

Выше мы говорили, что функции распределения безмассовых частиц  сохраняют
равновесную форму даже при выключении взаимодействия. Однако для нейтрино это не
так, даже в строго безмассовом пределе. Дело в следующем. При температурах, грубо говоря, выше 
МэВ, все частицы: фотоны, электроны, позитроны и нейтрино находятся в полном тепловом равновесии
и имеют одинаковые температуры. Нейтрино отключаются от электромагнитной компоненты плазмы,
т.е. от $\gamma, e^-, e^+$ при температуре ниже МэВ. Примерно в то же время при 
близких температурах происходит нагрев 
электромагнитной компоненты за счет $e^-e^+$-аннигиляции. Поэтому температуры нейтрино и 
электронов с позитронами станут различны. Посколько взаимодействие $e$ и $\nu$ выключается не
мгновенно, то обмен энергиями между горячими электронами и холодными нейтрино подогревает 
нейтрино, а так как $e\nu$-взаимодействие сильнее при высоких энергиях, то нейтрино нагреваются не
равномерно, а сильнее в высокоэнергичной части спектра, что и приводит к его неравновесной форме.
Искажение спектра было впервые посчитано в работе~\cite{ad-mf}, причем аналитически, и было
получено:
\be
\frac{\delta f_{\nu_e}}{ f_{\nu_e} }\approx 3\cdot 10^{-4} \,\,\frac{E}{ T}
\left( \frac{11 E }{ 4T } - 3\right)
\label{delta-f-nu}
\ee
(точнее в цитируемой работе был пропущен фактор 2 и поправка к спектру оказалась вдвое меньше
приведенного здесь выражения). 
Точные численные расчеты~\cite{dhs} подтверждают результат (\ref{delta-f-nu}), который
приводит к тому, что плотность энергии $\nu_e$ вырастет, 
как $\delta \rho_{\nu_e} /\rho_{\nu_e} = 0.83\%$. Поправки к спектрам $\nu_\mu$ и $\nu_\tau$ примерно вдвое меньше.  Соотношение с другими работами, где были проведены аналогичные вычисления,
обсуждается в обзоре~\cite{ad-nu-cosm}.  
В результате этого дополнительного нагрева суммарная плотность энергии всех трех типов нейтрино
будет эффективно отвечать не трем типам нейтрино, 
для которых нагрев не учтен, а 3.0339.  Кроме этого, плазменные 
поправки уменьшают плотность энергии фотонной компоненты~\cite{nu-plasma}, что еще немного
увеличивает {\it относительную} плотность энергии нейтрино, доводя ее до 3.0395.

Как известно, увеличение плотности энергии за счет дополнительных частиц по отношение к плотности 
фотонов в период первичного нуклеосинтеза влияет на выход легких элементов из-за зависимости темпа
охлаждения вселенной от числа типов различных частиц в плазме. Последнее дается коэффициентом
$g_*$ в уравнении (\ref{H-of-T}). В частности, частица с 
плотностью энергии равной энергии одного безмассового (или легкого, $m < O(MeV)$) нейтрино
увеличилa бы количество произведенного $^4 He$ примерно на 5\%. Наивно следовало бы ожидать,
что обсуждаемая здесь поправка привела бы к увеличению гелия на 0.2\%. Однако эффект оказывается
на уровне 0.01\%. Дело в том, что отклонение спeктра $\nu_e$ от равновесного непосредственно
влияет на отношение числа нейтронов к протонам за счет реакций 
\be
p+e^- \lrar n + \nu_e,\,\,\,  p + \bar \nu_e \lrar n + e^+.
\label{n-p-transform}
\ee
Избыток $\nu_e$ при высоких энергиях приводит к уменьшению
количества нейтронов при формировании легких ядер и оба эффекта почти сокращаются
(см. уравнение  (\ref{r-n-p}) в Приложении II и обсуждение после него).
Однако влияние дополнительных нейтрино на угловые флуктуации фона реликтовых фотонов не
подавлено и при точности (когда-то в будущем?) измерений на процентном уровне надо ожидать отличия 
эффективного числа типов нейтрино, определенного по количеству первичного гелия или дейтерия,
от того же количества, определенного по CMB, примерно на 0.04.

Заметим еще, что в силу ненулевой массы нейтрино их распределение заметно отличается от равновесного
в современной вселенной, когда импульс нейтрино стал меньше их масс. Оно имеет вид
\be
f_\nu = \frac{1}{ 1 + \exp [( p-\mu)/T ] }.
\label{f-nu-now}
\ee
Параметр $T$, входящий в это распределение, убывает при расширении, как $1/a$ и в настоящее время
примерно равен $2^o$ К, но, очевидно, не является 
температурой, т.к. в экспоненту входит не энергия, а  импульс.

Химический потенциал, $\mu$, характериризует разность между плотностями нейтрино и антинейтрино,
т.е. лептонную асимметрию, если $\mu \neq \bar \mu$.
При сохранении этой разности в сопутствующем объеме {\footnote{ Сопутствующим обьемом называется
объем, который расширяется вместе со вселенной, т.е. его величина пропорциональна $a^3$}} 
отношение $\xi = \mu/T$ остается постоянным
при расширении. Напомним, что если асимметрия между $\nu$ и $\bar \nu$ была генерирована, когда
нейтрино были в равновесии, то $\bar \mu = -\mu$. Разность между плотностями нейтрино и антинейтрино
дается выражением (\ref{n-barn}). Величина химического потенциала нейтрино плохо известна.
Самое сильное ограничение получается из первичного нуклеосинтеза. Плотность энергии нейтрино плюс
антинейтрино при $\mu \neq 0$ оказывается выше, чем при $\mu = 0$:
\be
\rho_\nu +\rho_{\bar\nu} &=&
{1\over 2\pi^2} \int_0^\infty dp p^3 \left[{1\over e^{p/T -\xi}+1} +
{1\over e^{p/T +\xi}+1}\right]  \nonumber \\
&=&{7\over 8}\,{\pi^2 T^4 \over 15}\left[1  +
{30\over 7}\,\left({\xi \over \pi}\right)^2 +
{15\over 7}\,\left({\xi \over \pi}\right)^4 \right].
\label{rhoxi}
\ee
Эта дополнительная энергия может быть описана, как возрастание эффективного числа нейтрино:
\be
\Delta N_\nu = {15\over 7}\left[ \left({\xi \over \pi}\right)^4 +
2 \left({\xi\over \pi}\right)^2 \right].
\label{deltanuxi}
\ee
Согласно современным данным первичный нуклеосинтез разрешает $\Delta N_\nu \approx 1$ и 
соответственно $\xi <1.5$. Это ограничение справедливо для $\nu_\mu$ и $\nu_\tau$. Для $\nu_e$
ограничение заметно сильнее: $\xi_{\nu_e} <0.1$, потому что, согласно сказанному выше, кинетика
реакции (\ref{n-p-transform}) напрямую зависит от распределения электронных нейтрино. В этот
момент уместно вспомнить об осцилляциях нейтрино. При экспериментально измеренных больших
углах перемешивания лептонная асимметрия в секторе любого типа нейтрино равномерно перемешивается
между всеми тремя типами, $\nu_e$, $\nu_\mu$  и $\nu_\tau$. Поэтому ограничение $\xi <0.1$ 
должно выполняться для любого типа нейтрино~\cite{dhpprs}.

\subsection{Закалка (замерзание) концентрации массивных частиц} \label{ss-freezing}

Довольно интересная кинетика проявляется при аннигиляции массивных частиц
(обозначим их $X$) в ранней вселенной.
Если бы равновесие поддерживалось ''до  бесконечности'', а асимметрия между этими частицами
отсутствовала, т.е. были бы равны концентрации таких частиц и античастиц, то к настоящему времени
их бы совсем не осталось и никакого вклада в темную материю они бы не сделали. Однако, аннгиляция
в какой-то момент останавливается и после этого количество этих частиц в сопутствующем 
обьеме становится постоянным. Приближенно можно оценить космологическую плотность этих
частиц следующим образом. Надо сравнить темп аннигиляции $\Gamma_{ann} =\sigma_{ann} v n_X$
с темпом космологического расширения, $H \sim T^2/m_{Pl}$. Аннигиляция эффективно прекратится,
когда станет $\Gamma_{ann} < H$. Отсюда, полагая $T\sim m_X$, получим, что плотность $X$- частиц
по отношению к плотности фотонов стремится к постояннoй величине:
\be
n_X/n_\gamma \approx \left( \sigma_{ann} vm_X m_{Pl} \right)^{-1}.
\label{nX-ngamma}
\ee 
Это выражение определяет, в частности, плотность частиц темной материи, если известно сечение
их взаимной аннигиляции. Точный результат отличается от приведенного лишь в несколько раз.

Аккуратный расчет асимптотической концентрации стабильных, но аннигилирующих $X$-частиц
требует небольших упрощающих предположений и простых манипуляций с кинетическим уравнением 
(\ref{lhs-cosm}) для $f_X$. Во-первых,  предполагается,
что в интеграл столкновений (\ref{I-coll}) доминирующий вклад вносит упругое рассеяние, которое
намного быстрее Хаббловского расширения, что приводит к функции распределения $X$-частиц вида:
\be
f_X = \exp \left[- (E_X- \mu)/T \right].
\label{f-X}
\ee
Здесь неявно содержится предположение, что замерзание $X$-частих происходит при $T\ll m_X$, 
когда становится справедливой статистика Больцмана. Параметр $\mu$ называют эффективным
химическим потенциалом и в случае равного числа $X$ и $\bar X$ частиц, т.е., как говорят 
зарядово-симметричной $X$-материи, химические потенциалы должны быть равны:
$\bar \mu = \mu$. Предполагаетеся, что в начальном состянии $\bar \mu =\mu =0$.
Возникновение ненулевых химических потенциалов 
отражает  выход из равнoвесия процесса аннигиляции.

Теперь мы сделаем простое, но замечательное действие, которое позволит исключить из интеграла
столкновений доминантную упругую часть, а оставить лишь аннигиляционную. Именно, проинтегрируем
обе части кинетического уравнения по $d^3 p_X/[2E_X (2\pi)^3]$. В результате большой вклад упругого 
рассеяния исчезает из интеграла столкновений, но тем не менее он 
фиксирует очень простую зависимость $f_X$ 
от энергии (\ref{f-X}). Оставшийся аннигиляционный интеграл легко берется и мы получим: 
\be
\dot n_X = -\langle \sigma_{ann} v \rangle \left( n^2_X - n_X^{(eq) 2}\right),
\label{n-X-dot}
\ee
где скобки $\langle ...\rangle$ означают усреднение по равновесному температурному распределению.
Для нерелятивистских частиц это усреднение несущественно.

Это уравнение было впервые введено в работах~\cite{zeld-eq} в 1965 году и было использавано в 
работах~\cite{lw-vdz} для вывода космологического ограничениня на массу тяжелого лептона, поэтому
это уравнение нередко называют уравнением Ли-Ваинберга, что не вполне справедливо.

В пределе большой величины $\sigma_{ann}v $ это уравнение с хорошей точностью решается аналитически,
 для чего интервал изменения $n_X$ делится на две части, когда $n_X \approx n_X^{(eq)} $ и когда
$n_X^{(eq)} \gg n_X$ и вторым слагаемым в скобках можно пренебречь. Описание такого, хотя и 
приближенного, но весьма точного аналитического решения приведено,
например, в обзоре~\cite{ad-yz}.

В последнее время довольно широко рассматривается возможность так называемой асимметричной
темной материи, когда на ранней космологической стадии возник избыток частиц над античастицами
(аналогично бариосинтезу). В этом случае античастицы почти совершенно исчезают, а плотность 
оставшихся частиц определяется их первоначальным избытком.

\section{Кинетика при нарушении СРТ} \label{s-CPT}

Как мы видели выше в разделе \ref{s-kineq}, 
нарушение С и СР не влияет на форму равновесных распределений частиц
и античастиц. Условие детального баланса, которое справедливо при наличии Т-инвариантности,
не выполняется, если  нарушено СР, хотя и имеет место СРТ теорема. Однако условия унитарности матрицы
рассеяния или СРТ-инвариантности при сохранении вероятности обеспечивает выполнение так
называемого циклического баланса, что в свою очередь приводит к сохранению канонической 
формы равновесных распределений (\ref{f-eq}). Однако, если СРТ-инвариантность разрушена, то
в зависимости от степени ''патологичности'' теории канонические равновесные распределения могут
измениться. Пока предположим, что этого не происходит. Возможность изменения равновесных
распределений при нарушении СРТ обсуждается ниже.

При нарушении СРТ массы частиц и античастиц могут стать разными.  Временно допустим, что это
действительно так. 
Если массы не равны, то плотности частиц и античастиц в тепловом равновесии при нулевых
химических потенциалах будут различны и, в частности, барионная асимметрия может возникнуть
в состоянии термодинамического равновесия:
\be{
n_B -n_{\bar B} \approx {g_s q_B \over 4\pi^2} {(m^2 - \bar m^2 ) \,T}
\int _1^\infty dy \, \exp (-my/T) \,\sqrt {y^2 -1},
\label{deltam}
}\ee
где $q_B$-барионный заряд рассматриваемых частиц. Однако для возникновения ненулевой плотности
барионного числа необходимо его несохранение на какой-то стадии. При сохранении
$B$ плотность барионного числа должна оставаться равной нулю и это приводит к генерации
ненулевого химического потенциала:
\be
\mu = \frac{\int (d^3 p/ E) f^2 (E,T,0) \,e^{E/T}  }{2 \int d^3 p f^2(E,T,0) \,e^{E/T}}\,
{m\delta m}.
\label{mu}
\ee

При неравенстве масс частиц и античастиц может оказаться возможной генерация барионной асимметрии
на электрослабой шкале несколько выше электрослабого фазового перехода. Как принято считать, при 
этих температурах не сохраняются ни барионное, ни лептонное числа. В стандартной теории этот сценарий
приводит к ничтожно малой барионной асимметрии из-за малого отклонения от 
равновесия и слабости СР-нарушения. Однако при нарушении 
СРТ и возникающего  в силу этого различия масс кварков и антикварков барионная асимметрия  	 		 
вполне может достигать наблюдаемого значения, т.к. не требуется нарушения термодинамического
равновесия. Вычисляя равновесные значения химических потенциалов, получим для барионного
числа в единице объема~\cite{ad-BG-CPT}:
\be
n_B = - \frac{T \ln 2 }{(2\pi)^3 }\left(\frac{9}{2} m_u \delta m_u +
\frac{15}{4} m_d\delta m_d \right).
\label{n-B}
\ee
Соответственно барионная асимметрия равна:
\be
\beta_T = \frac{n_B}{n_\gamma} = - 8.4\cdot 10^{-3} \left( 18 m_u\delta m_u + 
15 m_d \delta m_d \right) / T^2,
\label{beta}
\ee
где $n_\gamma = 0.24 T^3$ равновесная плотность числа фотонов в единице объема. Т.к. температура
должна быть выше температуры электрослабого перехода $T_{EW}$, то непренебрежимый вклад вносит 
самый тяжелый $t$-кварк. 

Выше предполагается неравенство масс частиц и античастиц, однако это вовсе не обязательно.
В работе~\cite{cdnt} приведен явный пример нелокальной Лоренц-инвариантной теории, в которй
нарушается СРТ, но сохраняется С и в силу этого массы частиц и античастиц должны быть равны.
Более того, при неравных массах теория предсказывает, что не будут сохраняться ни электрический, 
ни барионный заряды, ни даже тензор энергии-импулься~\cite{ad-vn}. По-видимому, в механизме
нарушения СРТ, рассмотренного в работе~\cite{cft}, этих проблем (или весьма интересных предсказаний?)
не возникает, но ценой весьма необычного дисперсионного закона, т.е. связи между энергией и импульсом, 
для свободных частиц. Если при неравных массах частиц и античастиц действительно не сохраняется
барионное число, то бариосинтез мог бы происходить и при температурах ниже $T_{EW}$. 

Как отмечалось в начале этого раздела, нарушение СРТ могло бы проявиться в отклонении равновесных
распределений от канонической формы (\ref{f-eq}). Строгого описания этого явления не существует, но
можно попытаться, следуя работе~\cite{ad-BG-CPT},  смоделировать его, предположив, что условие 
детального баланса нарушается, даже когда открыт лишь один канал реакции. Согласно обсуждению
в конце раздела~\ref{s-kineq}, это противоречит условию унитарности, но раз уж мы нарушили СРТ,  то не
исключено, что и унитарность тоже окажется нарушенной. Так что предположим, что амплитуды прямой
и обратной реакции различаются согласно соотношению:
\be
|A_{fi}|^2 = |A_{if}|^2 \left( 1 + \Delta_{if} \right).
\label{Delta-if}
\ee
Рассмотрим для определенности реакцию $a_1+a_2 \lrar a_3 + a_4$, полагая, что 
$a_j$-фермионы, хотя это и не обязательно. Предположим, что теперь равновесные функции распределения 
слегка отличаются от своих канонических значений, $f_j^{(eq)} = f_j(1+\delta_j)$, где
$f_j$-обычная равновесная функция (\ref{f-eq}). Подставив эти функции распределения в стандартное
кинетическое уравнение (хотя и его справедливость можно поставить под сомнение), найдем:
\be
f_1 \dot \delta_1 =\frac{1} {2E_1} \int \tilde{dp_2} \tilde{dp_3} \tilde{dp_4}  (2\pi)^4
\delta^{(4)} \left( p_1+p_2 -p_3-p_4\right) |A_{12}|^2
f_1 f_2 (1-f_3)(1-f_4) \nonumber \\
\left[ \Delta  +  \delta_3 \frac{2-f_3}{1-f_3} + \delta_4 \frac{2-f_4}{1-f_4}-
\delta_1 \frac{2-f_1}{1-f_1} -\delta_2 \frac{2-f_2}{1-f_2},
\right],
\label{dot-delta-f}
\ee
где $\tilde{dp} = d^3p/2E (2\pi)^2 $.

Равновесное состяние определяется условием того, что функция распределения не зависит от времени,
т.е. $\dot\delta_j = 0$. Однако фактор в квадратных скобках не может 
тождественно обратится в ноль, т.к. каждая
функция $f_j$ и $\delta_j$ зависит от ''своей''  энергии $E_j$. Зануление интеграла столкновений
возможно лишь в среднем после интегрирования по фазовому обьему. Более того, 
совершенно не очевидно, что решение $\dot \delta =0$ существует и, даже если это так, 
то форма равновесного распределения зависит от конкретной реакции, а 
универсальных равновесных распределений нет. Это сильно противоречит нашему привычному опыту.

\section{Приложение I: Термодинамические величины в равновесной плазме}  \label{P1}
 
Ниже приведены некоторые полезные выражения для интегральных характеристик 
равновесной слабовзаимодействующей плазмы или газа  в предположении, что 
функции распределения даются выражениями (\ref{f-eq}).

Сначала докажем общее соотношение между интегралами от бозонных и фермионных 
функций распределения для безмассовых частиц с нулевыми химическими 
потенциалами:
\be 
\int \frac{d^3p}{(2\pi)^3} \,p^n f_f = 
\left( 1 - \frac{1}{2^{2+n}}\right)\, \int \frac{d^3p}{(2\pi)^3} \,p^n f_b.
\label{P-f-b}
\ee
Введем безразмерную переменную $y = p/T$ и перепишем существенную часть
фермионного интеграла в виде:
\be
I_f = \int\,\frac{ dy\,y^{2+n}}{ e^y + 1 } =  
\int\ dy\,y^{2+n}\,\frac{e^y +1 -2}{ e^{2y} - 1 } =
\int\ dy\,y^{2+n}\, \left(\frac{1} {e^y -1 } - \frac{2}{ e^{2y} - 1 } \right).
\label{I-f}
\ee
Сделав во втором слагаемом замену $y \rar 2y $, получим искомое 
соотношение (\ref{P-f-b}).

Число частиц в единице объема и их плотность энергии даются соответственно 
интегралами (\ref{n}) и (\ref{rho}) и для безмассовых бозонов
(или при $T\gg m$) с нулевым химическим потенциалом они равны:
\be
n_b (0,T) &=& \frac{g_s \zeta (3) }{\pi^2}\, T^3 \approx 0.1218 g_s T^3,
\label{n-b-0} \\
\rho_b  (0,T) &=& \frac{g_s \pi^2}{30}\,T^4 ,
\label{rho-b-0}
\ee
где $\zeta (3) \approx 0.12$ - дзета функция. Для фермионов эти выражения надо
умножить соответственно на 3/4 и 7/8. 

В частности, для  фотонов микроволнового фона получим:
\be 
n_\gamma &=& 0.2436\,T^3=411.87 \,(T/2.728\, K)^3 \,\mbox{см}^{-3},
\label{n-CMB}\\
\rho_\gamma &=& 0.6215
\left({T}/{2.728{K}}\right)^4 {{ \mbox{эВ}}}/{{\mbox{см}}^3}.
\label{rho-CMB}
\ee

В случае массивных частиц в пределе $m\gg T$ для бозонов и фермионов получается
одинаковый результат:
\be 
n_{b,f} (m,T) = g_s \left(\frac{m T}{2\pi}\right)^{3/2}\,e^{-m/T}.
\label{n-b-f-m}
\ee
Очевидно, плотность энергии получается из этого выражения умножением на массу частиц: 
$ \rho (m,T) = mn(m,T)$.

\section{Приложение II: Примеры законов расширения и тепловая история 
вселенной} \label{P2}.

Эволюция масштабного фактора, т.е. зависимость $a(t)$, определяется космологическими уравнениями (\ref{fridmaneq1}) и  (\ref{dot-rho}) и
уравнением состояния материи (\ref{eq-of-state}). Легко видеть, что при $w=0$
плотность энергии убывает, как $\rho_{nr} \sim 1/a^3$, что совершенно очевидно
для нерелятивистского вещества. В этом случае $ a \sim t^{2/3}$ и $H = 2/3t$.
Для релятивистского вещества $w= 1/3$ и $\rho \sim 1/a^4$. Соответственно
$a \sim t^{1/2}$ и $H=1/2t$.  Еще один практически интересный случай - это 
уравнение состояния с $w=-1$. Оно реализовывалось на очень ранней стадии, инфляции, и 
приближенно, но с хорошей точностью реализуется сейчас. При этом
плотность энергии  и, следовательно, и параметр Хаббла остаются постоянными, 
а масштабный фактор растет экспонециально, $ a \sim \exp (Ht) $.

При движении вспять по времени температура вещества возрастает, грубо говоря, как
$T\sim 1/a(t)$. Более точно закон изменения температуры определяется 
уравнением (\ref{H-of-T}). На релятивистской стадии имеет место удобное 
приближенное соотношение
\be 
(t /{\rm sec})\,  (T/{\rm MeV})^{2} \approx 1.
\label{t-of-T}
\ee 
Мы уже отмечали, что на ранней стадии космологичекой эволюции первичная плазма
находилась практически в равновесном состоянии и все частицы с $m\geq T$ 
присутствовали в плазме примерно в равном количестве. 

В каком-то смысле вопреки вышесказанному, начальное состояние вселенной до
большого  взрыва было холодным. Оно представляло собой вакуумо-подобное
состояние скалярного поля-инфлатона $\phi$, которое приводило к экспоненциальному
расширению вселенной. Всякое другое вещество отсутствовало и говорить о 
температуре просто не было смысла. Согласно общепринятой картине, амплитуда 
инфлатонного поля медленно убывала и из-за этого убывал и параметр Хаббла.
Когда значение его упало до величины порядка массы инфлатона, то $\phi$ стало осциллировать
вокруг положения равновесия, эффективно рождая все элементарные частицы, масса 
которых была меньше массы инфлатона. Этот короткий период можно назвать большим
взрывом (Big Bang), с которого началась обычная тепловая космологическая история.
Частицы, порожденные инфлатоном, имели сильно неравновесный спектр, но за  очень
короткое время по сравнению с $1/H$, они термализовались и забыли начальные условия.

В какой-то момент после большого взрыва в результате процесса бариосинтеза
возник избыток частиц над античастицами. Где-то либо позже, либо раньше этого 
момента при охлаждении в 
первичной плазме происходили фазовые переходы, например,  переход от 
симметричной к несимметричной фазе в электрослабых взаимодействиях или
переход от фазы свободных кварков к фазе невылетания в квантовой 
хромодинамике. Когда температура упала примерно до 1 МэВ начались процессы
подготовки к первичному нуклеосинтезу, завершившиеся созданием легких элементов
при $T \approx 60-70$ кэВ.  Дальнейшее охлаждение проходило без особо ярких событий
примерно до $ z = 10^4 $. С этого момента плотность энергии нерелятивистского вещества
становится доминирующей и начинается рост первичных возмущений плотности, которые
впоследствии разовьются в крупномасштабную структуру вселенной (т.е. в галактики, их
скопления и сверхскопления). Несколько позже, при  $z=1100$, происходит рекомбинация 
водорода, т.е. протон захватывает электрон, образуя водород. Вещество становится нейтральным 
и фотоны, которые ранее лишь медленно диффундировали в заряженной плазме, начинают 
распространяться свободно, принося нам фотографию вселенной в то время.
Ниже мы остановимся подробнее на некоторых из упомянутых выше процессов.

Для того чтобы возник избыток барионов над 
антибарионами в исходно зарядово симметричной плазме, 
т.е. в плазме, которая состояла из равного числа частиц и античастиц достаточно выполнения
следующих трех условий~\cite{sakharov}: несохранения барионого числа, нарушения симметрии
взаимодействий частиц и античастиц, т.е. нарушения С и СР инвариантности и отклонения от
термодинамического равновесия. Детальное обсужение этих условий можно найти, например,
в лекциях~\cite{ad-varenna}. Отклонение от равновесия, как мы видели, всегда имеет место
для массивных частиц, см. уравнение (\ref{delta-f2}). Для того, чтобы эффект не был пренебрежимо мал,
масса частиц не должна быть меньше, чем $10^{-9} m_{Pl} \approx  10^{10}$ ГэВ. Поскольку
тяжелые частицы живут в плазме при $T$ выше или порядка $m$,  
то в этом случае бариосинтез может идти только
при очень высоких температурах или же масса Планка, $m_{Pl}$, в ранней вселенной должна быть
намного ниже своего каноничeского значения. Имеется, однако,
и другой механизм нарушения термодинамического равновесия в космологических условиях, а
именно фазовый переход первого рода. В процессе перехода вещество представляет собой смесь
двух различных фаз, как например, пар и вода в процессе кипения. Такое состояние явно неравновесно.
Подобные фазовые переходы могли бы произойти при $T \ll 10^{10}$ ГэВ, 
даже на масштабе КХД, но на
этом масштабе, скорее всего, сохраняется барионное число. К сожалению теория бариосинтеза,
хотя и описывает в принципе возникновение космологического избытка барионов над антибарионами,
но по существу представляет собой terra incognita, т.к. при необходимых для этого высоких 
температурах физика неизвестна и, скорее всего, не будет известна в обозримом будущем.

Напротив, первичный нуклеосинтез протекает в хорошо установленной области ядерной физики
низких энергий, поэтому в стандартной космологической модели предсказания теории однозначно 
определены. Сравнение теории с наблюдениями позволяет получить ограничения на новую физику, 
например, на существование неизвестных легких слабовзаимодействующих частиц или гипотетических
нестабильных, но долгоживущих частиц, или на свойства первичной плазмы, например, на величину
ее лептонной асимметрии. 
 
Исходным моментом первичного нуклесинтеза является фиксация протон-нейтронного отношения,
$r_{np}\equiv n_n/n_p$, что присходит в реакциях (\ref{n-p-transform}), идущих за счет слабого 
взаимодействия.  Темп этих реакций равен по порядку величины:
\be 
\Gamma_W \sim G_F^2 T^5,
\label{Gamma-W}
\ee
где $G_F \approx 10^{-5} $ ГэВ$^{-2}$ - константа Ферми слабого взаимодействия. Покуда скорость
этих реакций достаточно  высока, отношение $r_{np}$ следует равновесному закону:
\be
r_{np}= \exp \left(-\Delta m /T + \xi_{\nu_e} \right), 
\label{r-n-p}
\ee
где $\xi_{\nu_e}= \mu_{\nu_e}/T$, а $\mu_{\nu_e}$-химический потенциал электронных 
нейтрино. 

Сравнивая $\Gamma_W$ (\ref{Gamma-W})  с параметром Xаббла (\ref{H-of-T}), увидим, что 
реакции (\ref{n-p-transform}) замерзают при температуре, определяемой условием
$G_F^2 T_f^3 \sim  g_*T^2_f/m_{Pl} $, т.е. $ T_f \sim g_*^{1/3} $ МэВ. Более аккуратная оценка
приводит к коэффициенту порядка 0.16 
в этом выражении, так что при каноническом значении $g_* =10.75$
температура замерзания оказывается равной примерно 0.7 МэВ. В итоге отношение нейтронов к
протонам определяется выражением (\ref{r-n-p}), где надо подставить $T=T_f$. 
Отметим еще, что, например, увеличение силы взаимодействия $\nu_e$ приведет к уменьшению
$T_f$ и падению ''производства" гелия, т.к. 
почти все замерзшие нейтроны, за исключением их небольшой части (примерно, 20\%),
которая распадется к моменту образования ядер
при $T\sim 60-70$ кэВ, и тех нейтронов, что уйдут в дейтерий и гелий-3 (что составлет величину
около $10^{-4}$), соберутся в $^4 He$. Поэтому количество последнего весьма чувствительно
и к возможным дополнительным вкладам в плотность энергии (например, от новых частиц или
за счет ненулевых химических потенциалов нейтрино (\ref{rhoxi})). Особенно сильно количество
образовавшегося $^4 He$ зависит от химического потенциала $\nu_e$, как это видно из
выражения (\ref{r-n-p}).

Остановимся еще на сохранении энтропии при расширении. Можно показать, что в 
тепловом равновесии и при нулевых химических потенциалах сохраняется энтропия 
в сопутствующем объеме:
\be
\frac{dS}{dt} \equiv
\frac{d}{dt}\,\left(a^3\,\frac{\rho+P}{T}\right) = 0
\label{dS-dt}
\ee
где плотность энергии дается выражением (\ref{rho}), просуммированным по всем типам частиц,
''живущих'' в плазме, а плотность давления равна
\be
P = \sum \int_i \frac{d^3 q}{(2\pi)^3}\, \frac{q^2}{3E}\, f_i\left(\frac{E_i}{T}\right),
\label{P}
\ee
где сумма также берется по всем типам частиц.

В действительности равенство (\ref{dS-dt}) несколько более общее. Оно верно для 
любых, не обязательно равновесных, функции распределения для частицы типа $i$, 
которые имеют вид  $ f_i = f_i( E_i/T )$, 
с одинаковым параметром $T_i (t)$, 
и удовлетворяющих ковариантному закону эволюции плотности полной энергии:
${ \dot \rho = -3H \left(\rho + P\right)}$ с произвольной $T(t)$.
В результате легко получим
\be
{{ \frac{d}{dt}\,\left(a^3\,\frac{\rho+P}{T}\right)= }} 
{{ a^3\,\left[\frac{\rho+P}{T}\,\left(3H - \frac{\dot T}{T} - 3H\right)
+ \frac{\dot P}{T}\right].}
}\label{ds-dt2}
\ee
Используя выражение (\ref{P}) и учитывая, что в нем только ${ T}$ зависит от времени, найдем
после интегрирования  по частям:
\be
\dot P = \frac{\dot T}{T}\,(\rho + P)\,,
\label{dot-P}
\ee
что и приводит к закону сохранения (\ref{dS-dt}). Подчеркнем, что для справедливости этого закона
необходимо зануление химических потенциалов, что иногда упускается из вида. Отметим еще, что в 
литературе использутся не очень удачная терминология о выделении энтропии в распадах массивных
частиц или при их аннигиляции. Правильнее было бы говорить о перекачке энтропии от
массивных частиц к легким при $T<m$. В частности, из-за этого температура плазмы бесмассовых
частиц, которая при отсутствии аннигиляции или распадов падала бы, как $1/a$, будет падать
медленнее.  Например, аннигиляция  электронов и позитронов приводит к более высокой температуре 
фотонов по сравнению с нейтрино, температура которых падает, как $1/a$.

{\bf Благодарность.}  
Я выражаю признательность Е.В. Арбузовой, прочитавшей рукопись, за ценные замечания, а также  
благодарен за поддержку Гранту Правительства Российской Федерации 
 No. 11.G34.31.0047.

\end{document}